\documentclass[manuscript, acmsmall, nonacm]{acmart}
\setcopyright{none}

\usepackage{amsmath}
\usepackage{graphicx} 
\usepackage{subcaption} 
\usepackage{xcolor}

\usepackage{booktabs}   
\usepackage{tabularx}   
\usepackage{array}      
\usepackage{enumitem}   
\usepackage{ragged2e}   

\usepackage{etoolbox}
\AtBeginEnvironment{quote}{\itshape}

\AtBeginDocument{%
  }

\begin{document}

\title{A Guide to Using Social Media as a Geospatial Lens for Studying Public Opinion and Behavior}

\author{Lingyao Li}
\email{lingyaoli@arizona.edu}
\affiliation{%
  \institution{University of Arizona}
  \city{Tucson}
  \state{Arizona}
  \country{USA}
}

\renewcommand{\shortauthors}{Li et al.}

\begin{abstract}
\textbf{Abstract:} Social media and online review platforms have become valuable sources for studying how people express opinions, report experiences, and respond to events across space. This work presents a practical guide to using user-generated social data for geospatial research on public opinion, human behavior, and place-based experience. It shows the promise of using these data as a form of passive, distributed, and human-centered sensing that complements traditional surveys and sensor systems. Methodologically, the chapter outlines a general workflow that includes platform-aware data collection, information extraction, geospatial anchoring, and statistical modeling. It also discusses how advances in large language models (LLMs) strengthen the ability to extract structured information from noisy and unstructured content. Four case studies illustrate this framework: COVID-19 vaccine acceptance, earthquake damage assessment, airport service quality, and accessibility in urban environment. Across these cases, social media data are shown to support timely measurement of public attitudes, rapid approximation of geographically distributed impacts, and fine-grained understanding of place-based experiences.
\end{abstract}




\maketitle

\section{INTRODUCTION}

Understanding how people perceive risk and respond to unfolding events is central to public health, disaster management, and urban planning~\cite{bodas2022risk, lazer2020computational}. These questions matter not only for describing social and spatial conditions, but also for explaining how individuals and communities form opinions and act on them. Surveys and interviews have long been the primary tools for studying such processes~\cite{finnemann2024urban, han2024people, lazarus2021global, lazarus2023survey}, and they remain essential because they provide structured measurement and a strong basis for decision-making. However, they are often slow to deploy, costly to scale, and limited in their ability to capture rapidly evolving spatial and temporal dynamics.

Social media provides an important complementary lens~\cite{dinh2024social, li2025toward, li2022dynamic}. Platforms such as X/Twitter, Reddit, YouTube, and Google Maps host large volumes of user-generated content produced in direct response to lived experience. During the COVID-19 pandemic, users expressed views on vaccination~\cite{li2022dynamic}, masking~\cite{he2021people}, and reopening~\cite{li2021can}; during disasters, they reported urgent requests~\cite{zou2023social}, community response~\cite{ma2024investigating}, and infrastructure disruption~\cite{li2023exploring}; in everyday urban settings, online reviews capture experiences like parking~\cite{li2026crowdsourced} and service~\cite{kim2016impact} tied to specific places. These digital traces reveal how people interpret situations and signal behavioral intentions. Although imperfect proxies for broader public opinion, they offer an observational infrastructure that detects change rapidly and captures aspects of lived experience conventional surveys often miss~\cite{li2022dynamic, reveilhac2022systematic}.

For geospatial research, the value of these data lies in their ability to connect human expression to place~\cite{li2025toward, hasan2013understanding}. Some posts include explicit geographic coordinates, while others can be linked to location through profile metadata, place names in text, hashtags, images, or direct attachment to points of interest (POIs). This makes it possible to examine how attitudes, reported impacts, and everyday experiences vary across neighborhoods, cities, counties, states, and facilities. In addition, social media provides a form of passive data collection by recording what individuals choose to communicate in more natural settings~\cite{saha2019social}. These properties make social media well-suited for studying the spatial patterns of public opinion, human behavior, and place-based experience.

The analytical value of these data has expanded substantially with advances in natural language processing (NLP). Early work often relies on rule-based methods, sentiment lexicons, or conventional machine learning pipelines~\cite{hutto2014vader, li2022has, borg2020using, li2023exploring}. Transformer-based models such as Bidirectional Encoder Representations from Transformers (BERT) improve these approaches by enabling contextual text representations~\cite{devlin2019bert, kaliyar2021fakebert}. More recently, GPT-style large language models (LLMs) further change the landscape~\cite{lukito2024comparing, lukito2024comparing}. LLMs can extract structured information from noisy and highly unstructured text~\cite{duan2025crowdsourcing}, resolve references to entities and places~\cite{hu2023geo}, summarize large volumes of discussion~\cite{pereira2023crisis}, and synthesize evidence across multimodal inputs such as text and images~\cite{li2025pixels}. These capabilities are especially useful for social media analysis, where language is often informal, context-dependent, and highly variable across users and platforms~\cite{duan2025crowdsourcing}. In this sense, LLMs extend social media analytics beyond simple text classification toward richer forms of information extraction.

Against this background, this study conceptualizes social media as a form of crowdsourced geospatial sensing that is passive, distributed, and human-centered. The goal is not to position social media as a replacement for surveys or field inspection but as a complementary layer. Used in this way, social media can provide temporal immediacy and spatial detail that are often difficult to capture through conventional data sources alone. To illustrate this perspective, the study discusses common approaches for processing social media data and extracting information relevant to geospatial patterns, and then presents four case studies in different settings: (i) COVID-19 vaccine acceptance, (ii) earthquake damage assessment, (iii) airport service quality, and (iv) accessibility as a place-based urban experience. Together, these cases show how user-generated social data can extend geospatial research beyond locating events to understanding how people perceive, experience, and respond to them across space.
 
\section{METHODOLOGICAL FOUNDATIONS}
 
\begin{figure}[htbp]
  \centering
  \includegraphics[width=\linewidth]{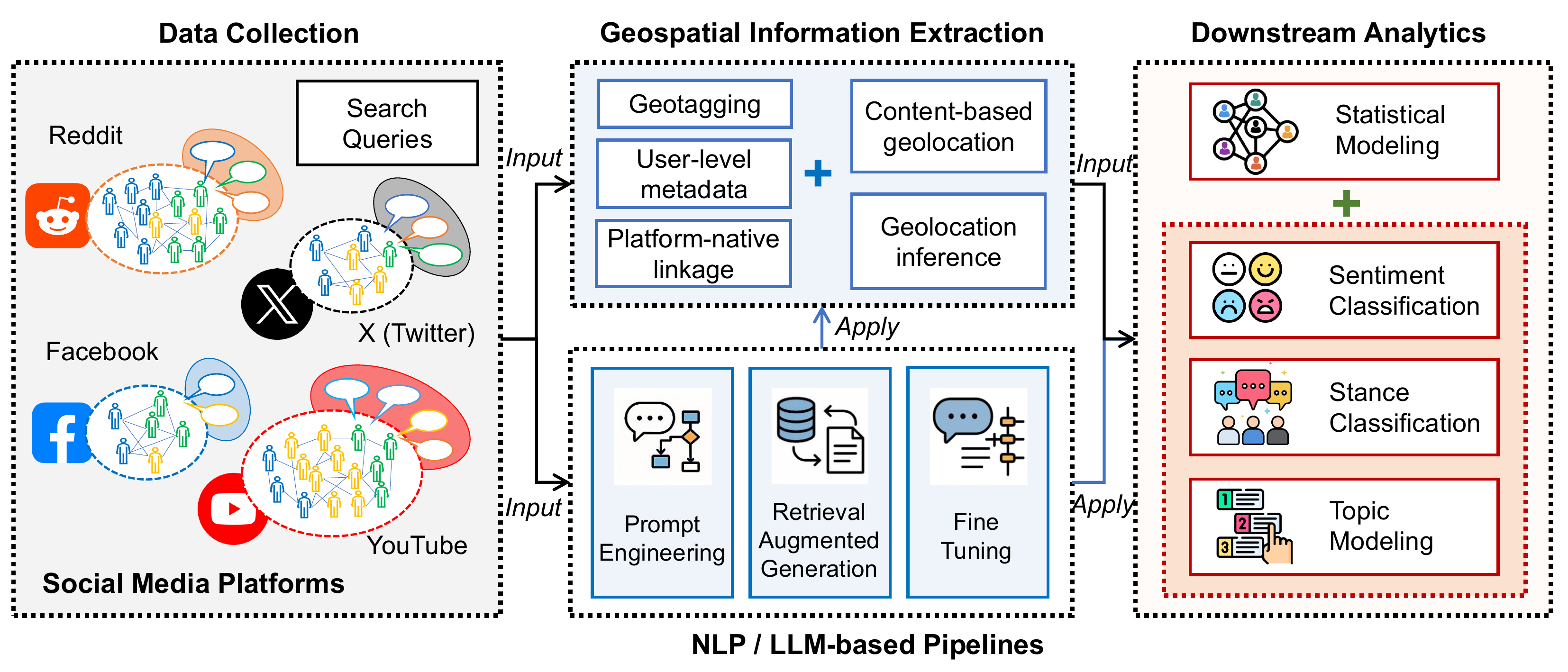}
  \caption{A general framework for processing social media data for geospatial analysis.}
  \label{fig:framework}
\end{figure}

Figure~\ref{fig:framework} outlines a general workflow for processing social media data for geospatial analysis. The pipeline begins with data collection from social media platforms, followed by computational models that can process textual, visual, and geographic information. The extracted signals are then linked to spatial units and analyzed using statistical models to identify geographic patterns, associations, and heterogeneity in public opinion and behavior. In practice, these stages are closely connected rather than strictly sequential: decisions made during data collection affect what can be extracted later, and the form of extracted information shapes the statistical inference. This section outlines the methodological foundations of this workflow, focusing on four core components: (i) data collection, (ii) information extraction, (iii) geospatial anchoring, and (iv) inferential modeling.
 
\subsection{Social Media Data Collection}

The first methodological decision is platform selection. Different platforms generate different forms of user expression and therefore support different research objectives, as summarized in Table~\ref{tab:platforms}. X (formerly Twitter) is characterized by rapid, event-driven posting and high temporal granularity, making it well-suited for tracking policy debate~\cite{milani2020visual}, crisis communication~\cite{cheng2018social}, and fast-changing public discourse~\cite{al2021public}. Reddit typically contains longer, more discursive posts useful for analyzing community interpretation, peer exchange, and extended discussion~\cite{gauthier2022will}. Facebook can capture localized coordination and civic communication~\cite{lappas2022harnessing}. Instagram and YouTube are particularly useful when visual content is central to the phenomenon under study~\cite{song2022analyze, mohamed2024users}. Google Maps and Yelp reviews differ because they are directly attached to POIs, enabling precise linkage to urban facilities and services~\cite{li2025toward, zhang2023can}.

\begin{table}[htbp]
  \centering
  \caption{Social media platforms and their geospatial analytic value.}
  \label{tab:platforms}
  \small
  \begin{tabular}{p{1.7cm} p{3.5cm} p{4.5cm} p{2.8cm}}
    \toprule
    \textbf{Platform} & \textbf{Content characteristics} & \textbf{Geospatial analytic value} & \textbf{Example studies} \\
    \midrule
    X (Twitter) & Short, rapid, event-driven posts with high temporal granularity & Tracking policy debate, crisis communication, and fast-changing public discourse & \citet{milani2020visual, cheng2018social, li2020leveraging} \\
    Reddit & Longer, discursive posts organized into topical communities & Analyzing community interpretation, peer exchange, and extended discussion & \citet{gauthier2022will, treen2022discussion} \\
    Facebook & Localized civic and community communication & Capturing neighborhood-level coordination and civic engagement & \citet{lappas2022harnessing} \\
    Instagram & Image-centric posts with captions and hashtags & Studying phenomena where visual content is central & \citet{song2022analyze} \\
    YouTube & Long-form video with comments and metadata & Multimodal analysis of events, opinions, and place-based experience & \citet{mohamed2024users} \\
    Google Maps & Reviews directly attached to POIs & Precise linkage of user evaluations to urban facilities and services & \citet{li2025toward} \\
    Yelp & POI-anchored reviews of businesses and services & Place-based service quality and consumer experience analysis & \citet{zhang2023can} \\
    \bottomrule
  \end{tabular}
\end{table}

Once the platform is selected, corpus construction requires explicit decisions about query design, time windows, language filters, deduplication rules, and inclusion criteria. Query design is particularly important because social media retrieval is intrinsically noisy. Narrow keyword rules can improve precision but may exclude relevant posts expressed through alternative wording, slang, abbreviations, or indirect references. Broader queries can improve recall but often introduce irrelevant content. For this reason, corpus construction typically requires iterative query refinement. In these workflows, an initial candidate set can be retrieved using keywords, hashtags, or Boolean rules, and then filtered with an LLM or a smaller domain-specific classifier to distinguish relevant posts from keyword matches.

Data engineering decisions also shape downstream analysis. Researchers need to decide whether to retain reposts, replies, and quoted posts; whether the unit of analysis should be the post, user, thread, or location; and whether reposts should be treated as evidence of prevalence or as a signal of diffusion. In public-opinion studies, duplicated or highly propagated content can bias prevalence estimates if aggregation is not handled carefully~\cite{hemphill2021drives}. In crisis communication studies, by contrast, reposting behavior may itself be analytically meaningful because it reflects information visibility and dissemination~\cite{xu2022analysing}.
 
\subsection{Text Parsing: From Classical NLP to LLM-based Pipelines}
 
The core computational task in social media analysis is to transform unstructured and context-dependent posts into structured information that can support subsequent inference. Common tasks include sentiment classification, stance detection, topic modeling, named entity recognition (NER), event extraction, summarization, and multimodal interpretation. These tasks can be implemented using classical NLP pipelines~\cite{manning2014stanford, camacho2022tweetnlp}, transformer models~\cite{devlin2019bert, kaliyar2021fakebert}, or LLM-based workflows~\cite{hu2023geo, lyu2025gpt}. The appropriate approach depends on the complexity of the target construct, the availability of annotated data, and the trade-offs among accuracy, interpretability, and scalability.

Classical pipelines usually begin with text normalization, tokenization, and vectorization. A common baseline is term frequency--inverse document frequency (TF-IDF)~\cite{salton1988term}, which remains effective for short-text classification because it captures discriminative lexical cues with limited modeling overhead:
\begin{equation}
  \text{TF-IDF}(t,d) = \text{tf}(t,d)\cdot \log\frac{N}{n_t},
\end{equation}
where $\text{tf}(t,d)$ denotes the frequency of term $t$ in document $d$, $N$ is the total number of documents, and $n_t$ is the number of documents containing term $t$. Traditional classifiers such as multinomial naive Bayes, logistic regression, support vector machines, and random forests can then be trained on the resulting feature matrix. Beyond TF-IDF, pre-transformer workflows also frequently use word representations such as Word2Vec~\cite{mikolov2013distributed}, GloVe~\cite{pennington2014glove}, and FastText~\cite{bojanowski2017enriching}, which encode semantic meaning in dense vector space. These embeddings can be paired with neural architectures such as convolutional neural network (CNN) classifiers to improve text classification.
 
Transformer models improve on sparse lexical features by encoding contextual semantics. Models such as BERT~\cite{devlin2019bert} and RoBERTa~\cite{liu2019roberta} generate dense token and sentence representations that support fine-tuned classification, similarity, and topic clustering. In a standard fine-tuning setting, a contextual encoder produces a representation $\mathbf{h}$ for an input post $x$, and the class probability is estimated as
\begin{equation}
  p(y\mid x) = \mathrm{softmax}(W\mathbf{h} + b),
\end{equation}
where $W$ and $b$ are trainable parameters. Contextual encoders are particularly important when meaning depends on context rather than isolated keywords. For example, a post can contain negative emotion (e.g., fear of COVID-19) while expressing support for the lockdown policy.
 
LLMs extend this progression in three main ways. First, they support instruction-based extraction, allowing researchers to specify a target schema directly in natural language. Second, they support few-shot learning~\cite{brown2020language} or fine-tuning~\cite{hu2022lora}, which is useful when labeled samples are small or discourse shifts rapidly over time. Third, LLMs can integrate textual and visual cues, enabling interpretation of captions, street scenes, and imagery~\cite{li2025pixels}. In abstract form, an LLM-based information extraction pipeline can be written as,
\begin{equation}
  y = \mathrm{LLM}(p, x, m),
\end{equation}
where $x$ is the social media post, $p$ is the instruction or prompt, $m$ denotes optional multimodal inputs such as images, and $y$ is the structured output. Although LLM-based pipelines offer substantial advantages over traditional methods, they can be computationally intensive and financially costly, particularly when models such as GPT or Gemini are applied to millions of social media posts.

\subsubsection{Sentiment Classification}
 
Sentiment classification estimates the polarity of a post as positive, negative, or neutral. Classical approaches often rely on rule-based or lexicon-based methods such as VADER~\cite{hutto2014vader}, which aggregate word-level polarity scores. These methods are computationally efficient and interpretable, but they are sensitive to slang, sarcasm, domain shift, and rapidly evolving platform-specific language. Supervised sentiment analysis replaces fixed lexical rules with learned mappings from text to labels. Typical workflows use TF-IDF features with classifiers such as logistic regression or support vector machines, whereas more recent workflows fine-tune contextual encoders such as BERT~\cite{gao2019target} or RoBERTa~\cite{barbieri2020tweeteval}. In either case, the objective is to learn a function $f$ that maps a post $x$ to a sentiment label $y$. Model performance is typically evaluated using precision, recall, and F1-score.
 
LLM-based sentiment analysis can be performed through prompt engineering. This is particularly useful when sentiment must be interpreted under informal or highly variable social media writing styles~\cite{zhang2024sentiment}. Compared with conventional classifiers, LLMs are often better able to interpret such language. They are also useful when sentiment must be extracted at a finer granularity than the whole post. In aspect-based sentiment analysis, for example, the goal is to estimate sentiment toward a specific attribute~\cite{nadi2023sentiment} (e.g., parking cost, accessibility, cleanliness, safety, or waiting time) from a Google Maps or Yelp review rather than assign a single polarity to the entire post.
 
\subsubsection{Stance Detection}
 
Stance detection estimates a post's orientation toward a specified target, such as a policy, intervention, public issue, or public figure~\cite{kuccuk2020stance}. It is distinct from sentiment analysis because emotional tone and target-oriented position do not necessarily coincide. A post may be emotionally negative while supporting a vaccine, or emotionally positive while opposing a government policy. Stance detection should therefore be treated as a separate inferential task rather than approximated by general sentiment.
 
Traditional stance detection typically relies on annotated datasets and supervised learning models using text vectorization and classifiers~\cite{li2022dynamic}. More recent approaches use contextual encoders such as BERT to model the relationship between the post and the target more explicitly~\cite{karande2021stance, li2026crowdsourced}. LLM-based stance detection further extends this capability by allowing the model to reason directly over the post and a specified target proposition~\cite{ziems2024can}. Instead of relying only on lexical polarity, LLMs can infer whether the post supports, opposes, or is unrelated to the target, and can optionally return supporting evidence or explanations.

An important extension is aspect-based stance detection, in which stance is estimated not only toward a broad issue but toward a specific dimension of that issue. For example, a post may support vaccination overall while opposing mandates or expressing concern about side effects. A related task is aspect-based sentiment classification~\cite{mughal2024comparative}, which estimates sentiment toward a specific attribute rather than the entire post. These finer-grained approaches are increasingly important because many public policy debates are multidimensional rather than binary.

\subsubsection{Topic Modeling}

When discourse is heterogeneous and the objective is exploratory, topic modeling can be used to identify latent thematic structure. Traditional latent Dirichlet allocation (LDA) represents each document as a mixture of topics and each topic as a distribution over words~\cite{jelodar2019latent}. Under this formulation, the probability of observing a word \(w\) in document \(d\) is,
\begin{equation}
  p(w \mid d) = \sum_{k=1}^{K} p(w \mid z = k)\, p(z = k \mid d),
\end{equation}
where \(z\) denotes the latent topic assignment and \(K\) is the total number of topics. LDA remains interpretable, but its bag-of-words assumption is often limiting for short social media posts, where meaning depends heavily on context and co-occurrence patterns are sparse.
 
Recent topic modeling workflows increasingly rely on BERTopic~\cite{grootendorst2022bertopic}, which is better suited to context-dependent text. BERTopic first computes transformer-based document embeddings, then applies dimensionality reduction, typically with Uniform Manifold Approximation and Projection (UMAP)~\cite{mcinnes2018umap}, to preserve semantic structure in a lower-dimensional space. The reduced embeddings are then clustered, often using HDBSCAN, and each cluster is summarized using class-based TF--IDF to identify representative terms. This combination of semantic embeddings, dimensionality reduction, density-based clustering, and lexical summarization makes BERTopic more effective than traditional bag-of-words topic models for many social media datasets.

LLMs can further enhance topic analysis in two ways. First, they can improve cluster interpretability by generating clearer and more coherent human-readable labels. Second, they can summarize groups of posts into analytic themes while preserving representative evidence or example posts. This is especially useful for social media discourse analysis, where a single cluster may contain eyewitness reports, advice, emotional reactions, and media commentary. Recent GPT-based workflows are best understood as LLM-assisted topic interpretation rather than replacements for embedding-based clustering. One example is TopicGPT, a prompt-based topic modeling framework~\cite{pham2024topicgpt}.

\subsubsection{Named Entity Recognition (NER) and Event Extraction}

NER identifies spans corresponding to entities such as persons, organizations, places, and facilities~\cite{li2020survey}. In geospatial social sensing, NER often provides the first bridge between text and location because explicit geographic coordinates are uncommon. Classical systems such as Stanford CoreNLP~\cite{manning2014stanford} and other neural sequence-labeling tools~\cite{hemati2019lstmvoter} have been widely used for this task, particularly when the target entity set is known in advance. However, extracting place names alone is usually insufficient. Event extraction must also determine what happened, where it happened, and in what status or stage. For example, a wildfire post may mention a city while also indicating whether it concerns an active evacuation, a shelter destination, or general discussion.

LLM-based pipelines can help simplify this process by enabling direct extraction into structured information. For example, an LLM can be prompted to return fields such as event type, stage, location, stance classification, and supporting evidence in a JSON schema~\cite{li2025llm}. This joint extraction strategy is especially useful for noisy social media text because it can integrate entity recognition and contextual interpretation in a single step. When possible, structured extraction should retain evidence spans or source sentences so that the output remains auditable.

\subsection{Geospatial Anchoring and Location Inference}
 
\begin{figure}[htbp]
  \centering
  \includegraphics[width=0.9\linewidth]{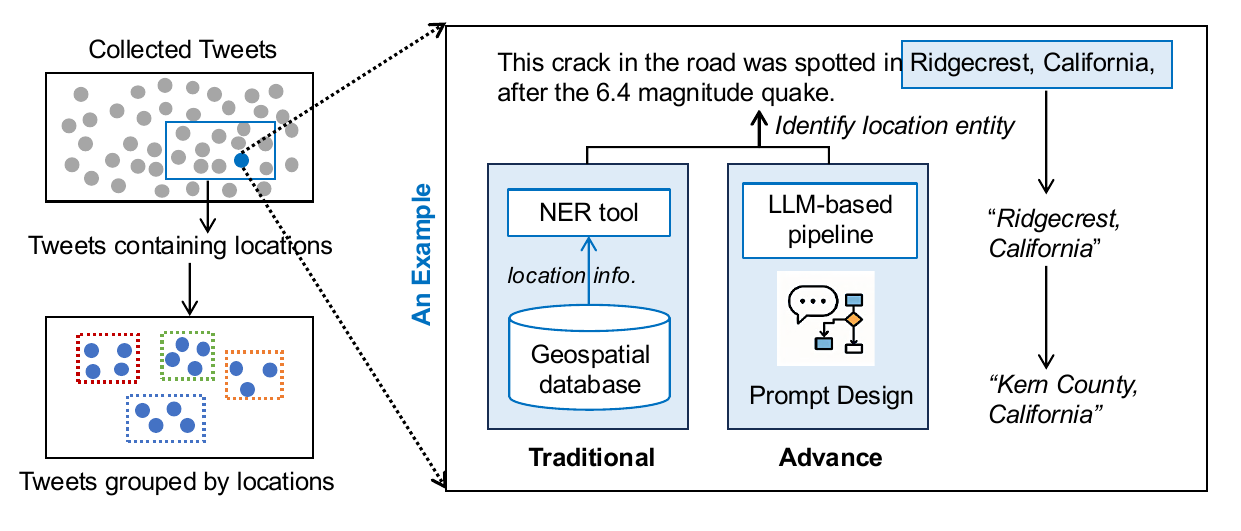}
  \caption{An example of content-based geolocation information extraction.}
  \label{fig:location}
\end{figure}

The geospatial value of social media depends on how reliably digital traces can be linked to places. In most workflows, location is not directly observed but inferred from multiple signals with different levels of spatial precision and uncertainty. Several strategies are commonly used for location inference, as summarized in Table~\ref{tab:geoanchor}. The first is direct geotagging, in which latitude-longitude coordinates are attached to the post itself. This provides the highest spatial precision, but geotagged posts are rare on most platforms. The second is user-level metadata, such as profile location or home region, which increases coverage but often reflects the user's general location rather than the location of the reported event or experience. The third is content-based geolocation (illustrated in Figure~\ref{fig:location}), in which location is inferred from text or hashtags using NER. The fourth is platform-native linkage, as in Google Maps or Yelp reviews, where content is already attached to a POI and therefore inherits a natural spatial anchor. The fifth is geolocation inference using LLMs, which infer location from joint textual and visual context, such as road signs, architectural style, terrain, storefronts, or other scene-level cues. This strategy is particularly useful when explicit coordinates and place names are absent but visual evidence is available.

\begin{table}[htbp]
  \centering
  \caption{Common strategies for geospatial anchoring in social media analysis.}
  \label{tab:geoanchor}
  \small
  \begin{tabular}{p{2cm} p{3.9cm} p{4.7cm} p{1.8cm}}
    \toprule
    \textbf{Strategy} & \textbf{Description} & \textbf{Strengths and limitations} & \textbf{Example studies} \\
    \midrule
    Direct geotagging & Latitude--longitude coordinates attached to the post itself & Highest spatial precision, but geotagged posts are rare on most platforms & \citet{paradkar2022examining} \\
    
    User-level metadata & Profile location or self-reported home region & Broader coverage, but reflects general user location rather than the event location & \citet{li2022dynamic} \\
    
    Content-based geolocation & Place names, hashtags, and textual cues extracted via NER & Wide applicability, but sensitive to ambiguity, nesting, and multiple references & \citet{li2021data} \\
    
    Platform-native linkage & Content already attached to a POI (e.g., Google Maps, Yelp) & Natural and reliable spatial anchor, limited to review-style platforms & \citet{li2026crowdsourced} \\
    
    LLM-based inference & Joint interpretation of textual and visual cues by multimodal LLMs & Useful when coordinates and place names are absent but visual evidence is available; uncertainty in inference & \citet{li2025pixels} \\
    \bottomrule
  \end{tabular}
\end{table}

For geolocation inference, content-based geolocation is a common approach (see Figure~\ref{fig:location}). Classical NLP pipelines typically use NER tools such as Stanford CoreNLP~\cite{manning2014stanford} or spaCy~\cite{honnibal2020spacy} to identify candidate place mentions in text (e.g., ``Ridgecrest, California'' from the given tweet). In parallel, computer vision methods can infer location from images through landmark recognition, scene classification, or reverse geocoding when distinctive environmental cues are visible~\cite{hu2022beyond}. More recently, LLMs have been used to process images, allowing models to interpret place references, understand contextual descriptions, and infer likely locations based on visual information~\cite{li2025pixels}.
 
However, geolocation inference remains challenging. Place names may be ambiguous. A post may mention ``Springfield,'' which cannot be uniquely identified without additional context, or use relational expressions such as ``northern California'' or ``near Shaver Lake'' rather than naming a single canonical location. Posts may also contain multiple place references, some of which correspond to origins, destinations, comparison points, or broader administrative regions rather than the actual event location~\cite{li2021social}. Reliable geospatial inference therefore requires careful interpretation of which location is most relevant to the event or experience under study. For real-world applications, neighborhood-level precision can be difficult to validate; aggregation at the city, county, corridor, or metropolitan level often provides a more defensible balance between spatial coverage and locational certainty~\cite{li2025toward}.
 
\subsection{Statistical and Inferential Modeling}
 
Once social media posts have been transformed into analytic variables and linked to place, the next step is statistical modeling. This usually requires aggregation, validation, and inference. Aggregation is necessary because raw posts rarely correspond directly to the substantive unit of interest. Depending on the research question, indicators may be aggregated to users, locations, time periods, events, or POIs. The choice of aggregation unit should follow the theoretical construct being measured. Post-level aggregation emphasizes communication intensity and temporal responsiveness, whereas user-level aggregation reduces distortion from highly active accounts. Place-level aggregation supports spatial analysis, but estimates may become unstable in sparsely observed areas. Let $S_u$ denote an aggregated digital indicator for a unit $u$, such as mean sentiment, stance prevalence, or impact severity score. A general aggregation can be written as,
\begin{equation}
  S_u = \frac{1}{n_u}\sum_{i \in u} s_i,
\end{equation}
where $s_i$ is the extracted signal from post $i$ assigned to unit $u$, and $n_u$ is the number of posts or users contributing to that unit.
 
Validation is critical because social media indicators are only useful if they correspond, at least partially, to external phenomena of scientific interest. A common strategy is to compare derived indicators with downstream ground-truth data, such as vaccination rates, hazard intensity maps, or administrative statistics from nationwide surveys. This can be assessed through correlation analysis. For two variables $x$ and $y$ observed over $n$ spatial or temporal units, the Pearson correlation coefficient is~\cite{schober2018correlation},
\begin{equation}
  r = \frac{\sum_{i=1}^{n}(x_i - \bar{x})(y_i - \bar{y})}{\sqrt{\sum_{i=1}^{n}(x_i - \bar{x})^2 \sum_{i=1}^{n}(y_i - \bar{y})^2}}.
\end{equation}
A strong correlation does not prove causal validity, but it provides evidence that the extracted digital signal captures meaningful variation related to the target process~\cite{li2021social}. In longitudinal settings, validation can also be framed through time-series analysis or event-study logic, asking whether the indicator responds as expected to major external events~\cite{li2022dynamic}.
 
Regression modeling is then used to explain how social media--derived indicators vary with socioeconomic, demographic, environmental, or built-environment factors across geographic regions. A standard starting point is ordinary least squares (OLS)~\cite{draper1998applied}, written as,
\begin{equation}
  y_i = \beta_0 + \sum_{k=1}^{K} \beta_k x_{ik} + \varepsilon_i,
\end{equation}
where $y_i$ is the digital indicator for unit $i$, $x_{ik}$ are explanatory variables, $\beta_k$ are regression coefficients, and $\varepsilon_i$ is the error term. In this setting, regression analysis is useful for testing whether online public responses are systematically associated with contextual factors such as income, racial composition, education, land use, accessibility, or transportation infrastructure.

In spatial settings, however, OLS may be insufficient because relationships can vary across space and residuals may be spatially auto-correlated. Spatial regression models are therefore often more appropriate. Multiscale geographically weighted regression (MGWR)~\cite{fotheringham2017multiscale} allows coefficients to vary locally, thereby capturing geographic heterogeneity in the association between digital indicators and contextual variables. The Spatial Durbin Model (SDM)~\cite{anselin2022spatial} extends this by accounting for dependence between neighboring units. A general SDM can be written as,
\begin{equation}
  \mathbf{y} = \rho W\mathbf{y} + X\boldsymbol{\beta} + WX\boldsymbol{\theta} + \boldsymbol{\varepsilon},
\end{equation}
where $W$ is a spatial weights matrix, $\rho$ captures spatial dependence in the dependent variable, and $\boldsymbol{\theta}$ captures spillover effects from neighboring covariates. These models are particularly useful when public opinion or behavioral response is shaped not only by local conditions but also by social and spatial contexts.

\section{CASE STUDIES}

The four cases below are organized to illustrate how different computational tasks map onto different questions. The vaccine case emphasizes stance detection and index construction. The earthquake case illustrates damage-oriented text classification and location inference. The airport case investigates place-based service evaluation and shows how online reviews support geographically explicit comparison. The accessibility case highlights how user-generated reviews, combined with regression modeling, can reveal socio-spatial inequality in inclusive urban design.
 
\subsection{CASE 1: Vaccine Acceptance}
 
The vaccine case demonstrated how social media could be used not only to characterize online discourse but also to construct a geographically explicit indicator of public readiness for vaccination~\cite{li2022dynamic}. The study analyzed approximately 29 million English-language tweets containing the keyword \emph{vaccine} from August 2020 to April 2021. From this corpus, 15{,}000 frequently occurring unique tweets were manually labeled as positive, negative, or unrelated, and an additional 2{,}000 tweets were reserved for testing. Multiple text-classification pipelines were evaluated, and TF-IDF combined with a random forest classifier achieved the best overall performance with approximately 74.4\% testing accuracy. The selected model was then applied to classify the full corpus.

A major contribution of the study was the development of the Vaccine Acceptance Index (VAI). Rather than aggregating classified tweets directly, the study first computed a user-level acceptance score based on the relative balance of positive and negative tweets posted by each individual, and then averaged these values within each geographic unit. This design reduced the influence of highly active accounts and yielded a measure that was more interpretable across places and time. The VAI was estimated at multiple geographic scales, including the national, state, and county levels.
 
\begin{figure}[htbp]
  \centering
  \includegraphics[width=\linewidth]{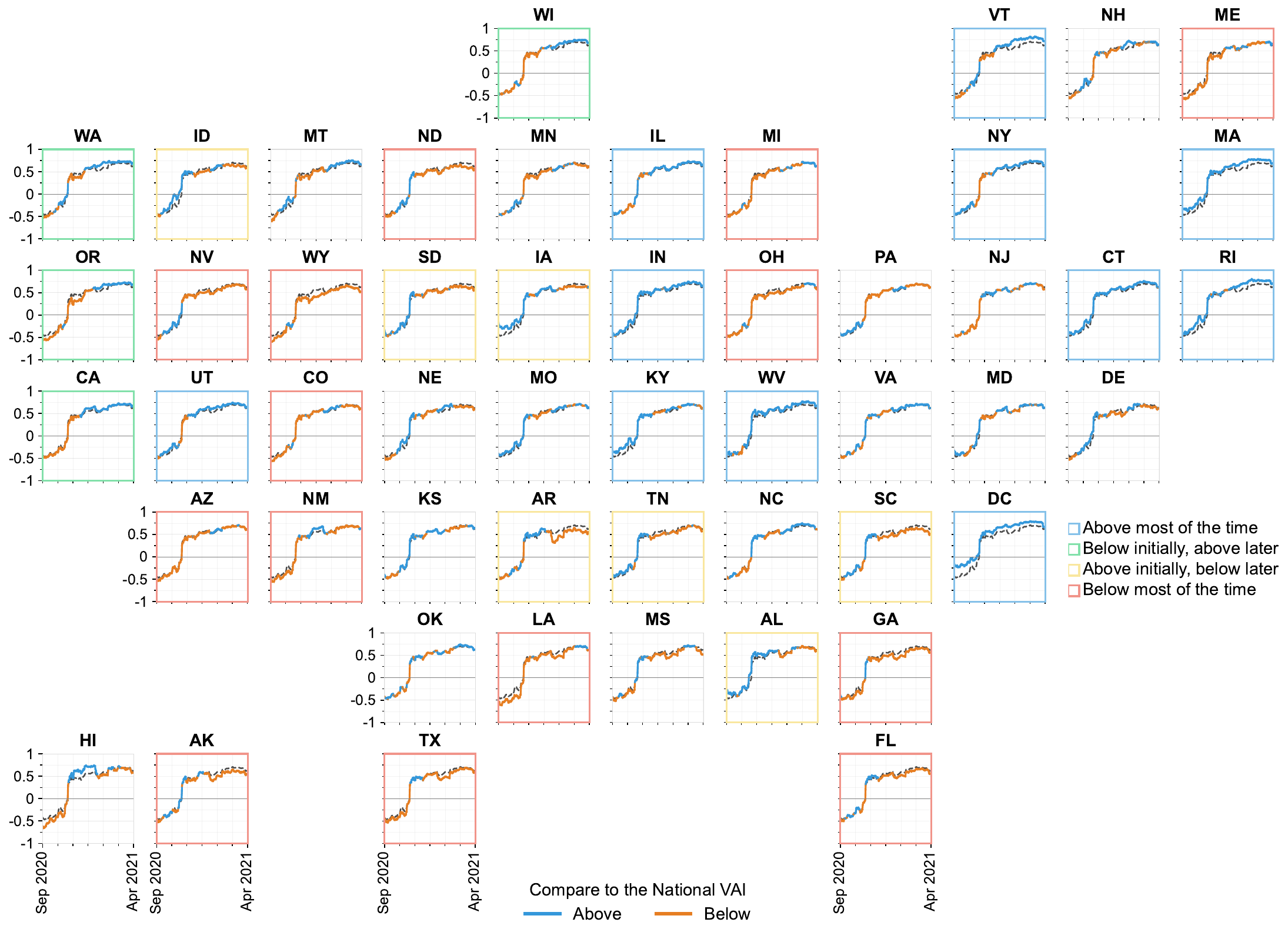}
  \caption{State-level trajectories of the VAI from August 2020 to April 2021. Each panel shows how a state's VAI changed over time relative to the national VAI (adjusted from~\cite{li2022dynamic}).}
  \label{fig:vaccine}
\end{figure}

Figure~\ref{fig:vaccine} showed how the VAI changed across states over time. At the national level, the index shifted from negative to positive in late 2020 and remained largely positive after January 2021, indicating a broad change in public orientation as vaccine development progressed and rollout expanded. The state-level results revealed substantial geographic heterogeneity in vaccine acceptance. The VAI could be mapped across states to show where acceptance was relatively higher or lower at different time points, thereby transforming diffuse online discussion into a spatially interpretable indicator of public opinion. For example, states such as New York and Massachusetts remained above the national VAI for much of the study period, whereas Texas and Florida were generally below it. California became more positive as rollout progressed, while some southern states showed weaker or declining acceptance after November 2020.
 
Another strength of this case was that the VAI was evaluated against external reference data rather than presented as an internally coherent metric alone. The study compared the index with subsequent vaccination uptake and with survey-based measures of vaccine hesitancy, including the CDC Household Pulse Survey~\cite{census_hps} and its county-level downscaling based on American Community Survey~\cite{census_acs} data. The results showed meaningful associations at both the state and county levels, with stronger relationships in counties that met a modest threshold of active users.

\subsection{CASE 2: Earthquake Damage Assessment}
 
The earthquake damage case addressed a key challenge in disaster response: obtaining an early picture of where damage was likely to be concentrated before formal assessment is complete~\cite{li2021social}. Using the 2019 Ridgecrest earthquake sequence as a case study, this analysis began with a large corpus of earthquake-related tweets and then filtered for damage-related content. A subset of tweets was manually labeled using a four-level damage scale derived from the Modified Mercalli Intensity framework: no damage, slight damage, moderate damage, and severe damage. Multiple text-classification pipelines were evaluated.

Location inference was equally important in this case. Damage reports were useful only when they could be tied to a place, yet most social media posts did not include precise geographic coordinates. The study therefore combined explicit coordinates, when available, with content-based location inference from textual references to assign damage levels to counties. This required distinguishing locations directly associated with reported damage from places mentioned only as comparison points, prior events, or general discussion. The case thus shows that rapid damage assessment depends not only on text classification, but also on geospatial anchoring.
 
\begin{figure}[htbp]
  \centering
  \includegraphics[width=\linewidth]{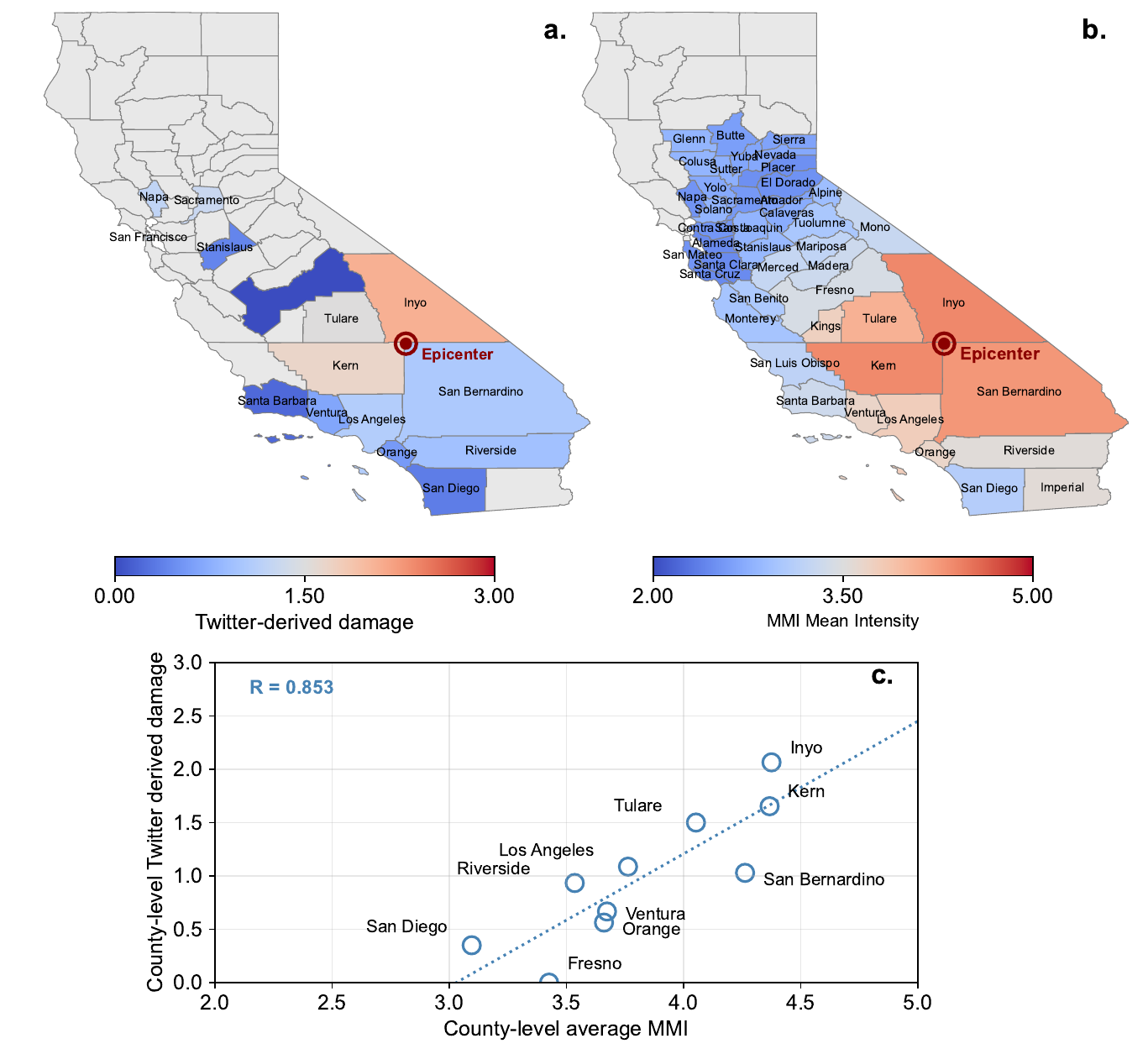}
  \caption{County-level earthquake damage assessment from Twitter compared with U.S. Geological Survey intensity data. (a) Twitter-derived county damage; (b) county-level average Modified Mercalli Intensity from the USGS \emph{Did You Feel It?} system; (c) the correlation between these two data resources, indicating that crowdsourced reports can approximate the spatial distribution of earthquake impact (adjusted from~\cite{li2021social}).}
  \label{fig:earthquake}
\end{figure}

Figure~\ref{fig:earthquake} presented the county-level results. The Twitter-derived county damage map is shown together with the U.S. Geological Survey \emph{Did You Feel It?} map to compare the spatial distribution of crowdsourced reports with the official communal intensity surface. The highest estimated damage levels were observed in counties near the epicenter, especially Kern and Inyo, whereas counties farther away showed substantially lower estimated damage. The study also showed that the tweet-derived damage signal increased rapidly after the major shocks and generally converged within approximately 12 to 14 hours, indicating that social media can provide a useful early approximation of damage geography.
 
At the same time, the study made clear that this approach should be interpreted as a rapid and approximate complement to conventional inspection-based systems rather than a replacement. This case further showed that social media users can function as distributed sensors whose reports, when carefully classified and geolocated, supplement traditional damage assessment with temporal immediacy and spatial density.
 
\subsection{CASE 3: Airport Service Quality}

The airport service quality case illustrated how online reviews could be used to measure place-based experience at scale~\cite{li2022has}. Google Maps reviews provide a complementary source of user-generated evaluations that are directly tied to specific airports. In this study, reviews from the 98 busiest U.S. airports were used to examine how passenger perceptions changed before and after the COVID-19 outbreak. The study combined topic extraction with fine-grained sentiment analysis. A topical ontology was developed to identify eight first-level service dimensions: access, check-in, security, wayfinding, arrival, facilities, environment, and personnel. Because each review could contain evaluations of multiple service dimensions, the study applied aspect-based sentiment analysis to estimate sentiment toward specific aspects of the airport experience rather than assigning a single overall polarity to the entire review.

\begin{figure}[htbp]
  \centering
  \includegraphics[width=\linewidth]{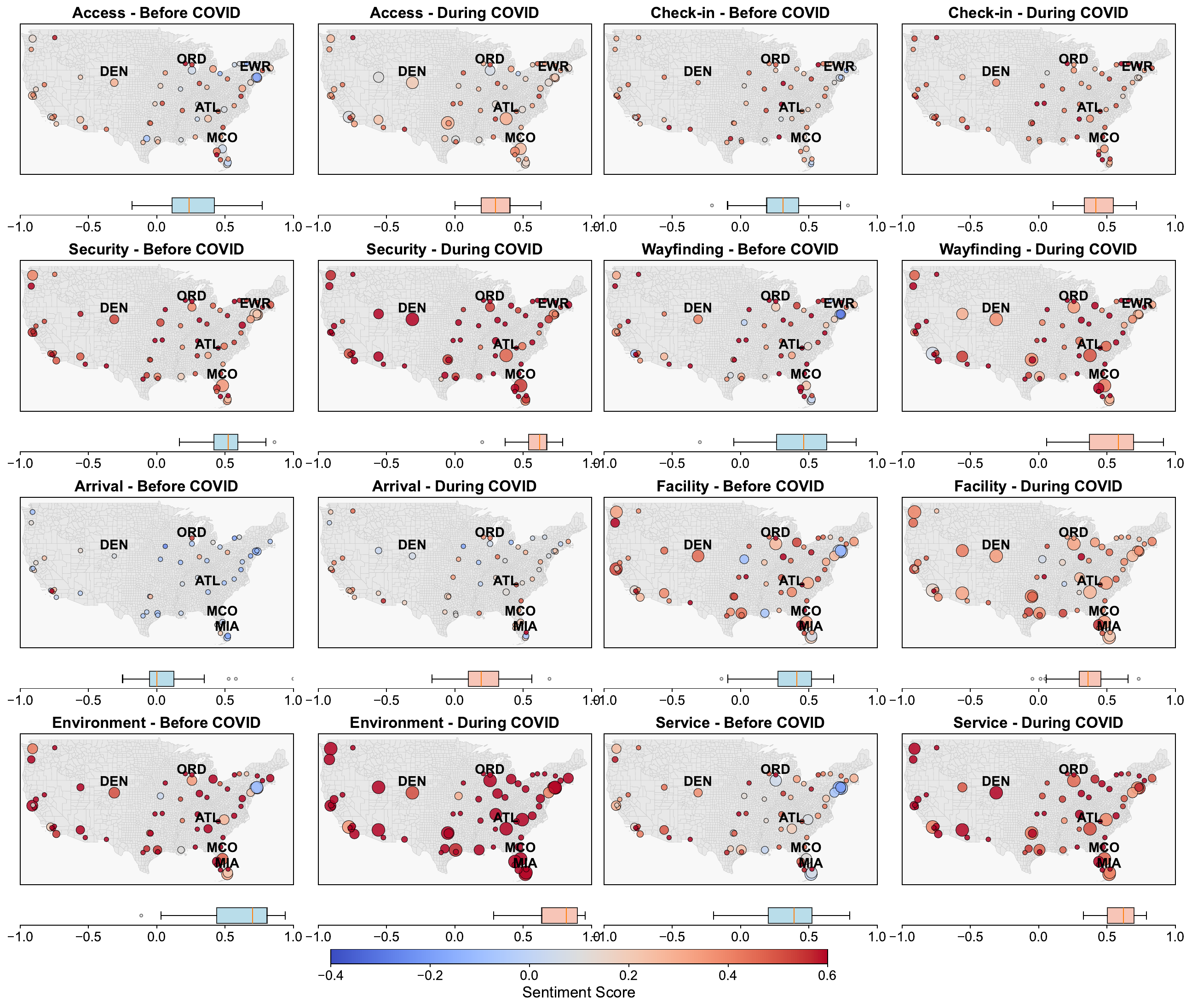}
  \caption{Airport-level sentiment maps for eight dimensions of airport service quality before and during COVID-19. Each panel shows the spatial distribution of topic-specific sentiment across the top 98 international U.S. airports (adjusted from~\cite{li2022has}).}
  \label{fig:airport}
\end{figure}

Figure~\ref{fig:airport} showed the airport-level sentiment before and during COVID-19. The results revealed clear temporal shifts after the COVID-19 outbreak. The average airport rating increased from 3.55 to 4.13, and sentiment improved for most service dimensions, especially personnel, environment, arrival, check-in, and wayfinding. Among these, personnel showed the largest increase in sentiment, while facilities remained largely unchanged. Environment had the highest average sentiment after the outbreak, suggesting that travelers responded positively to cleanliness and environmental conditions, whereas arrival remained the lowest-rated dimension in both periods, likely reflecting dissatisfaction with baggage claim and passport control processes.
 
The airport-level map revealed substantial heterogeneity across airports and service dimensions (Figure~\ref{fig:airport}). For example, the study noted that Atlanta airport (ATL in Figure~\ref{fig:airport}) showed less positive sentiment for environment before COVID-19 but improved substantially afterward. The map also illustrated that airport service quality was spatially uneven and multi-dimensional, showing that online reviews could function as a geographically explicit indicator of traveler experience.

\subsection{CASE 4: Accessibility Satisfaction}

The accessibility case shows how online reviews can be used to study inclusive urban design at a national scale~\cite{li2026crowdsourced}. Google Maps reviews provide a useful complementary source because they capture naturally occurring public experiences tied to specific POIs. In this study, more than one million accessibility-related reviews from POIs across the United States were used to investigate how people perceive accessibility in everyday urban environments. The study framed the problem as a specialized attitude-classification task rather than relying on generic sentiment tools.
 
\begin{figure}[htbp]
  \centering
  \includegraphics[width=\linewidth]{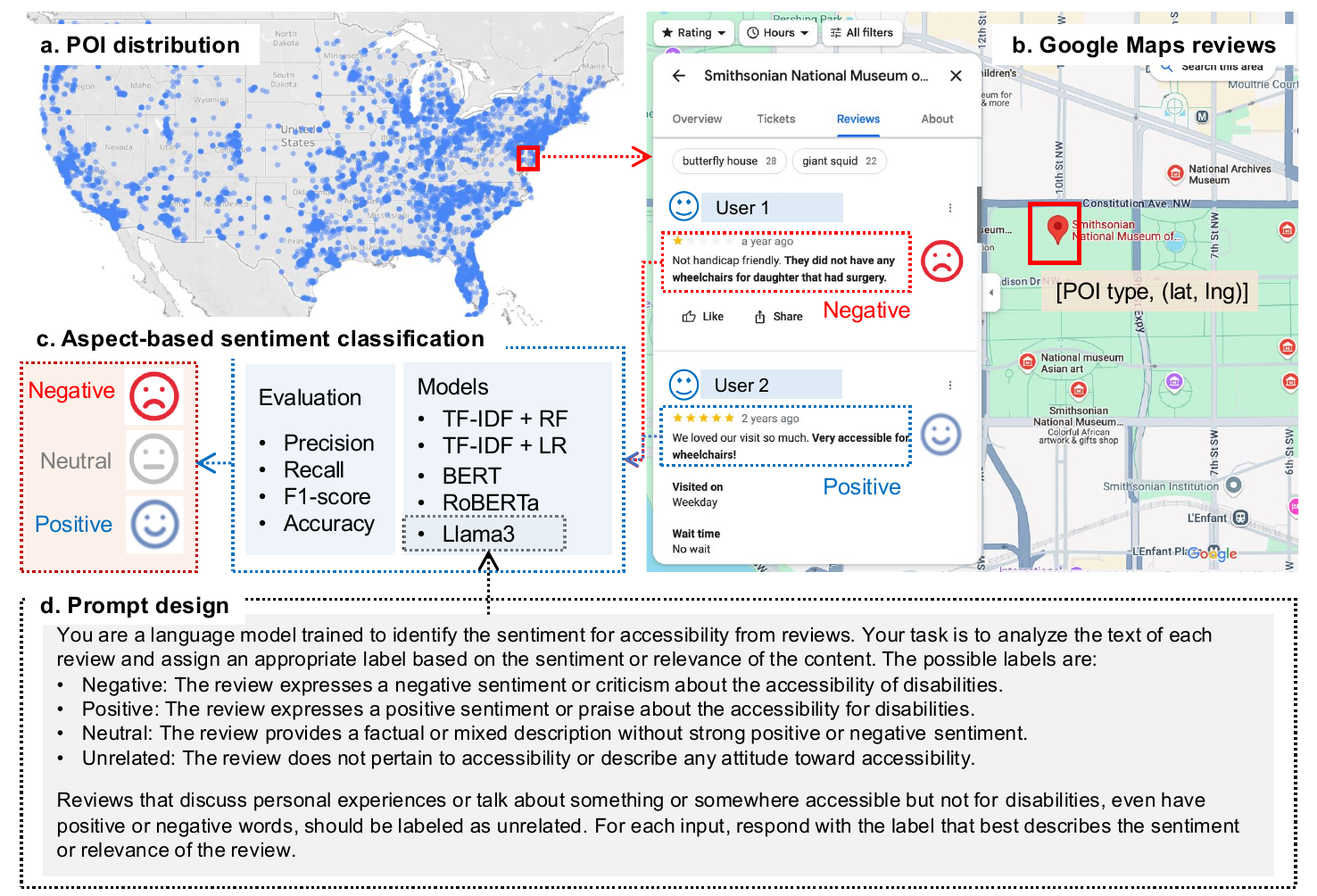}
  \caption{Workflow for extracting accessibility attitudes from Google Maps reviews. (a) The POI distribution; (b) A representative example of relevant Google Maps review; (c) Aspect-based sentiment classification; (d) Prompt design for LLM-based attitude classification (adjusted from~\cite{li2026crowdsourced}).}
  \label{fig:accessibility_workflow}
\end{figure}

The workflow (Figure~\ref{fig:accessibility_workflow}) began by constructing an accessibility-focused dictionary grounded in ADA guidelines and empirical review language, and then annotating reviews into positive, negative, neutral, or unrelated classes. A prompt design is illustrated in Figure~\ref{fig:accessibility_workflow}(d). After comparing multiple candidate models, the fine-tuned Llama 3 model achieved the strongest performance and was applied to the full dataset. This design allowed the analysis to move beyond keyword matching and recover more meaningful public attitudes toward accessible facilities and services.
 
\begin{figure}[htbp]
  \centering
  \includegraphics[width=\linewidth]{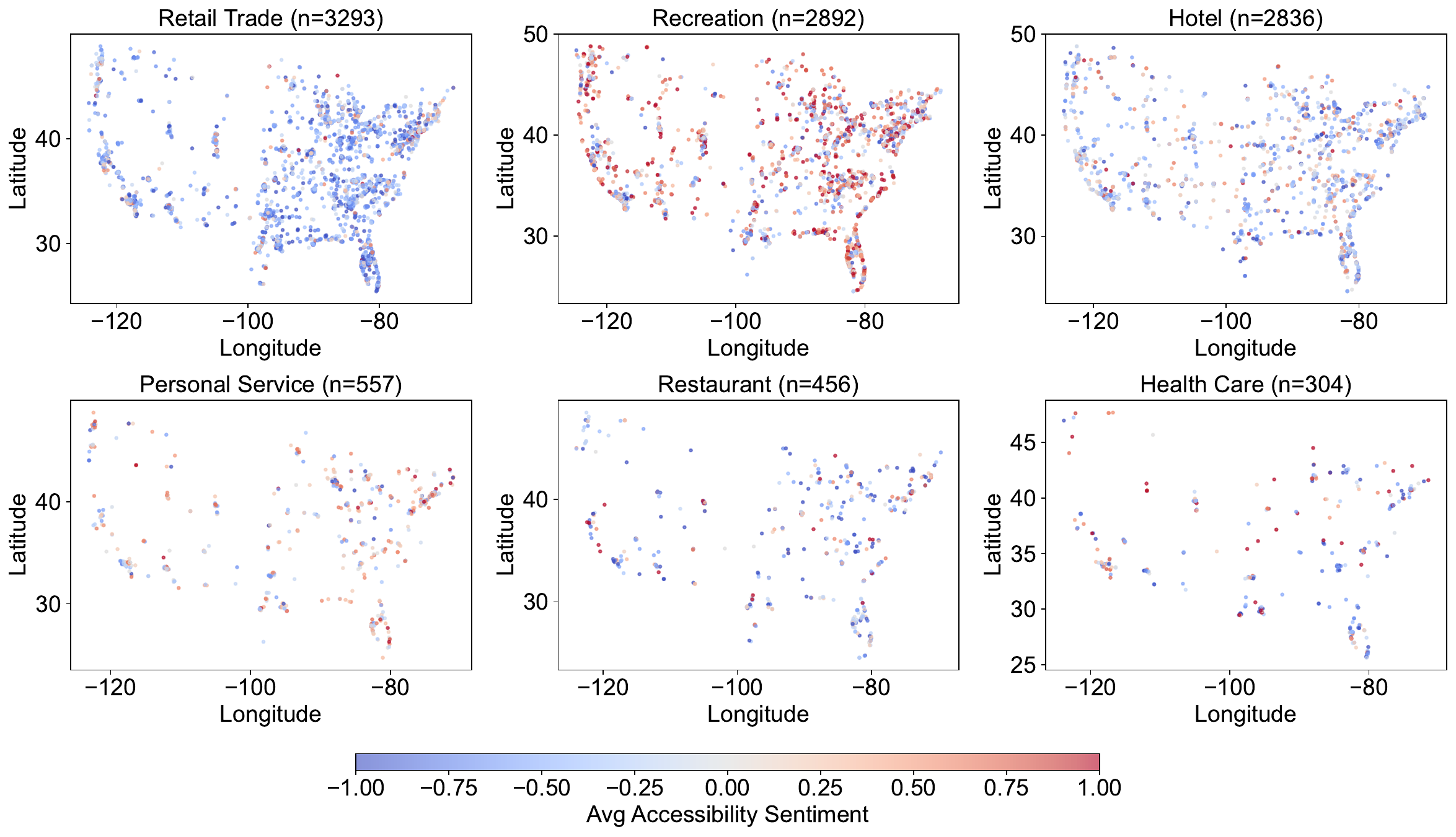}
  \caption{Average accessibility sentiment across POI categories and its spatial distribution in the U.S. The figure summarizes POI-level accessibility sentiment, including retail, recreation, hotel, personal service, restaurant, and health care.}
  \label{fig:accessibility_results}
\end{figure}

At the POI level, the results (Figure~\ref{fig:accessibility_results}) revealed substantial heterogeneity across urban activity spaces. Most major POI categories, including restaurants, retail, hotels, and health care, showed predominantly negative accessibility sentiment, suggesting persistent barriers across sectors that are central to daily life. By aggregating accessibility sentiment to counties and census block groups, the study examined how public perceptions varied across socio-spatial contexts. The regression results showed that more positive accessibility sentiment was associated with areas with higher proportions of white residents and greater socioeconomic advantage, whereas more negative sentiment was observed in areas with higher concentrations of elderly and highly educated populations. No clear relationship was found between disability prevalence and sentiment itself, but a significant positive association was identified between public sentiment and external disability-friendly scores.
 
These findings suggested that accessibility was not only a design issue at the level of individual facilities, but also a socio-spatially uneven urban condition. More broadly, the case showed how user-generated reviews could help planners and policymakers identify where accessibility challenges arose and how they related to broader patterns of urban inequality.

\section{DISCUSSION}
 
\subsection{Potential and Practical Implications}
 
One of the main strengths of social media for geospatial research is its temporal responsiveness. Because user-generated content is produced continuously and often in direct response to unfolding events, it can provide near-real-time evidence of public attitudes and place-based experiences. This is particularly valuable in settings where conventional data sources are too slow to support timely understanding, such as public health emergencies~\cite{kostkova2014swineflu} and disaster response~\cite{wu2018disaster}. In these contexts, social media is best understood not as a replacement for formal data systems, but as an early observational layer that can reveal emerging spatial patterns before surveys or administrative reporting become available.
 
A second strength lies in geographic scale. A single collection and processing pipeline can gather information across many cities, counties, states, or POIs simultaneously, making it possible to compare spatial heterogeneity in public opinion and behavior at scales that are difficult to achieve through conventional fieldwork alone~\cite{li2021can, li2026crowdsourced}. The case studies in this study illustrate this advantage clearly: vaccine acceptance can be tracked across states and counties~\cite{li2022dynamic}, and disaster impacts~\cite{li2021social} can be approximated across affected regions. In this sense, social media expands geospatial analysis from isolated case observation to scalable comparative measurement.

Social media also contributes a form of informational richness that is difficult to recover from conventional administrative data or physical sensing systems. User-generated content can show how people respond to events such as satisfaction, inconvenience, trust, fear, or behavioral intention. Combined with computational text analysis, these expressions can be translated into indicators of stance~\cite{li2022dynamic}, perceived impact~\cite{ma2024investigating}, service quality~\cite{kim2016impact}, or urban environmental barriers~\cite{li2026crowdsourced}. This makes it possible to examine subjective experience in relation to spatial context in a more systematic and analytically tractable way.
 
These characteristics also create substantial practical value for decision-making. For planners, transportation agencies, emergency managers, and public health officials, social media analytics can help identify where problems are concentrated, which issues matter most to the public, and how responses differ across places. In disaster settings, this may support rapid situational awareness~\cite{ma2023appraising}; in public health, it may help track shifts in confidence or hesitancy~\cite{li2022dynamic}; in urban planning, it may reveal persistent dissatisfaction with accessibility, parking, or service environments~\cite{li2026crowdsourced, li2025toward}. 

The broader implication is therefore methodological as well as substantive. The value of social media does not lie simply in the volume of available data, but in the ability to connect computational extraction with geospatial reasoning and statistical inference. The most informative applications are not those that merely count posts or classify sentiment, but those that link extracted signals to meaningful spatial units and interpret them in relation to demographic, socioeconomic, or built-environment conditions. Recent advances in LLMs and multimodal systems further strengthen this potential by making it easier to extract structured information from noisy text, interpret nuanced attitudes, resolve place references, summarize large corpora, and integrate textual and visual evidence. Social media is thus becoming not only a source of rapid descriptive signals, but also a foundation for richer forms of geospatial social sensing.
 
\subsection{Limitations and Future Work}
 
Understanding the limitations of social media data is equally important. The first is representativeness. Social media platform demographics differ by age, education, political engagement, urbanization, and digital access~\cite{blank2017representativeness, yin2018evaluating}. Even within a platform, those who post are not the same as those who read silently, and those who post repeatedly can dominate the visible signal. This means that social media indicators should not be interpreted as direct estimates of population prevalence without calibration or careful caveats.
 
A second limitation concerns geospatial precision. Explicit geotags are rare, profile locations are often coarse or outdated, and content-based place extraction is inherently noisy. A post may mention multiple places, relational geography, or locations unrelated to the focal event~\cite{li2021social}. For this reason, scientific claims should be aligned with the defensible level of spatial certainty. In many applications, city- or county-level inference is more credible than point-level mapping. Advanced LLMs can improve contextual interpretation, but they cannot recover geographic certainty that is absent from the source material itself.
 
A third limitation concerns the difficulty of extracting meaningful opinions from social media text. Posts are often unstructured and heavily shaped by platform-specific language, which makes interpretation inherently uncertain. In particular, constructs such as stance are difficult to identify because opinions are often implicit or mixed. A post may express fear, sarcasm, or frustration without clearly indicating support or opposition, and the same wording can carry different meanings depending on the context and timing. Similar challenges arise when trying to infer event status or behavioral intention from brief and noisy content. Although LLM-based pipelines improve the ability to interpret context and recover structured information, they also introduce additional uncertainty, including sensitivity to prompt design~\cite{zhuo2024prosa} and occasional hallucination~\cite{li2023halueval}. Scientific use therefore requires carefully designed pipelines and systematic human validation.
 
A fourth limitation is platform dependence. Data access policies, API pricing, and content moderation rules can change abruptly. A pipeline that is viable on one platform or during one period may not remain viable later, which directly affects reproducibility and the comparability of longitudinal research. For example, after Elon Musk acquired Twitter in October 2022, the company announced in February 2023 that free API access would be discontinued~\cite{dang2023x_api}, showing how platform-level policy shifts can quickly alter research feasibility. Future work should therefore emphasize portable methods that can operate across platforms, as well as archives that preserve legally usable research corpora when possible.
 
Looking ahead, several directions appear especially promising. First, modern LLMs can integrate text with images, maps, and street-level scenes, creating new opportunities for disaster impact assessment, environmental monitoring, and place-based urban analysis. Second, retrieval-augmented workflows can ground social media extraction in official or external data sources, thereby reducing unsupported inference and improving verifiability. Third, structured extraction can move beyond simple labels toward richer records that capture locations, time windows, infrastructure types, and confidence scores. Fourth, cross-platform data fusion may help reduce platform-specific bias by combining fast-moving streams such as X/Twitter with place-linked platforms such as Google Maps and more discursive environments such as Reddit.

Finally, ethical practice must remain central. Social media data may be public, but this does not make every analytical use ethically unproblematic. Researchers should minimize harm, avoid unjustified inference about individuals, respect platform terms and privacy norms, and remain cautious when analyzing sensitive behaviors or vulnerable communities. The scientific value of crowdsourced geospatial sensing therefore depends not only on computational sophistication, but also on transparent, responsible, and context-aware practice.

\bibliographystyle{ACM-Reference-Format}
\bibliography{references}

@article{lazarus2023survey,
  title={A survey of COVID-19 vaccine acceptance across 23 countries in 2022},
  author={Lazarus, Jeffrey V and Wyka, Katarzyna and White, Trenton M and Picchio, Camila A and Gostin, Lawrence O and Larson, Heidi J and Rabin, Kenneth and Ratzan, Scott C and Kamarulzaman, Adeeba and El-Mohandes, Ayman},
  journal={Nature medicine},
  volume={29},
  number={2},
  pages={366--375},
  year={2023},
  publisher={Nature Publishing Group US New York}
}

@article{lazarus2021global,
  title={A global survey of potential acceptance of a COVID-19 vaccine},
  author={Lazarus, Jeffrey V and Ratzan, Scott C and Palayew, Adam and Gostin, Lawrence O and Larson, Heidi J and Rabin, Kenneth and Kimball, Spencer and El-Mohandes, Ayman},
  journal={Nature medicine},
  volume={27},
  number={2},
  pages={225--228},
  year={2021},
  publisher={Nature Publishing Group US New York}
}

@article{han2024people,
  title={Why do people not prepare for disasters? A national survey from China},
  author={Han, Ziqiang and Wu, Guochun},
  journal={Npj Natural Hazards},
  volume={1},
  number={1},
  pages={1},
  year={2024},
  publisher={Nature Publishing Group UK London}
}

@article{finnemann2024urban,
  title={The urban desirability paradox: UK urban-rural differences in well-being, social satisfaction, and economic satisfaction},
  author={Finnemann, Adam and Huth, Karoline and Borsboom, Denny and Epskamp, Sacha and van der Maas, Han},
  journal={Science Advances},
  volume={10},
  number={29},
  pages={eadn1636},
  year={2024},
  publisher={American Association for the Advancement of Science}
}

@article{bodas2022risk,
  title={Risk perception of natural and human-made disasters—cross sectional study in eight countries in Europe and beyond},
  author={Bodas, Moran and Peleg, Kobi and Stolero, Nathan and Adini, Bruria},
  journal={Frontiers in public health},
  volume={10},
  pages={825985},
  year={2022},
  publisher={Frontiers Media SA}
}

@article{lazer2020computational,
  title={Computational social science: Obstacles and opportunities},
  author={Lazer, David MJ and Pentland, Alex and Watts, Duncan J and Aral, Sinan and Athey, Susan and Contractor, Noshir and Freelon, Deen and Gonzalez-Bailon, Sandra and King, Gary and Margetts, Helen and others},
  journal={Science},
  volume={369},
  number={6507},
  pages={1060--1062},
  year={2020},
  publisher={American Association for the Advancement of Science}
}

@article{li2021can,
  title={Can social media data be used to evaluate the risk of human interactions during the COVID-19 pandemic?},
  author={Li, Lingyao and Ma, Zihui and Lee, Hyesoo and Lee, Sanggyu},
  journal={International Journal of Disaster Risk Reduction},
  volume={56},
  pages={102142},
  year={2021},
  publisher={Elsevier}
}

@article{li2021data,
  title={Data-driven investigations of using social media to aid evacuations amid Western United States wildfire season},
  author={Li, Lingyao and Ma, Zihui and Cao, Tao},
  journal={Fire Safety Journal},
  volume={126},
  pages={103480},
  year={2021},
  publisher={Elsevier}
}

@article{li2022dynamic,
  title={Dynamic assessment of the COVID-19 vaccine acceptance leveraging social media data},
  author={Li, Lingyao and Zhou, Jiayan and Ma, Zihui and Bensi, Michelle T and Hall, Molly A and Baecher, Gregory B},
  journal={Journal of Biomedical Informatics},
  volume={129},
  pages={104054},
  year={2022},
  publisher={Elsevier}
}

@article{dinh2024social,
  title={Social Media and Crisis Informatics Research in LIS},
  author={Dinh, Ly and Hong, Lingzi and Dumas, Catherine and Patin, Beth and Ghosh, Souvick and Li, Lingyao and Khoury, Christy},
  journal={Proceedings of the Association for Information Science and Technology},
  volume={61},
  number={1},
  pages={749--753},
  year={2024},
  publisher={Wiley Online Library}
}

@article{ma2024investigating,
  title={Investigating disaster response for resilient communities through social media data and the Susceptible-Infected-Recovered (SIR) model: A case study of 2020 Western US wildfire season},
  author={Ma, Zihui and Li, Lingyao and Hemphill, Libby and Baecher, Gregory B and Yuan, Yubai},
  journal={Sustainable Cities and Society},
  volume={106},
  pages={105362},
  year={2024},
  publisher={Elsevier}
}

@article{li2021social,
  title={Social media crowdsourcing for rapid damage assessment following a sudden-onset natural hazard event},
  author={Li, Lingyao and Bensi, Michelle and Cui, Qingbin and Baecher, Gregory B and Huang, You},
  journal={International Journal of Information Management},
  volume={60},
  pages={102378},
  year={2021},
  publisher={Elsevier}
}

@article{li2023exploring,
  title={Exploring the potential of social media crowdsourcing for post-earthquake damage assessment},
  author={Li, Lingyao and Bensi, Michelle and Baecher, Gregory},
  journal={International Journal of Disaster Risk Reduction},
  volume={98},
  pages={104062},
  year={2023},
  publisher={Elsevier}
}

@article{li2026crowdsourced,
  title={Crowdsourced reviews reveal substantial disparities in public perceptions of parking},
  author={Li, Lingyao and Hu, Songhua and Dinh, Ly and Hemphill, Libby},
  journal={Cities},
  volume={171},
  pages={106866},
  year={2026},
  publisher={Elsevier}
}

@article{li2025toward,
  title={Toward satisfactory public accessibility: A crowdsourcing approach through online reviews to inclusive urban design},
  author={Li, Lingyao and Hu, Songhua and Dai, Yinpei and Deng, Min and Momeni, Parisa and Laverghetta, Gabriel and Fan, Lizhou and Ma, Zihui and Wang, Xi and Ma, Siyuan and others},
  journal={Computers, Environment and Urban Systems},
  volume={122},
  pages={102329},
  year={2025},
  publisher={Elsevier}
}

@article{reveilhac2022systematic,
  title={A systematic literature review of how and whether social media data can complement traditional survey data to study public opinion},
  author={Reveilhac, Maud and Steinmetz, Stephanie and Morselli, Davide},
  journal={Multimedia tools and applications},
  volume={81},
  number={7},
  pages={10107--10142},
  year={2022},
  publisher={Springer}
}

@article{he2021people,
  title={Why do people oppose mask wearing? A comprehensive analysis of US tweets during the COVID-19 pandemic},
  author={He, Lu and He, Changyang and Reynolds, Tera L and Bai, Qiushi and Huang, Yicong and Li, Chen and Zheng, Kai and Chen, Yunan},
  journal={Journal of the American Medical Informatics Association},
  volume={28},
  number={7},
  pages={1564--1573},
  year={2021},
  publisher={Oxford University Press}
}

@article{kim2016impact,
  title={The impact of social media reviews on restaurant performance: The moderating role of excellence certificate},
  author={Kim, Woo Gon and Li, Jun Justin and Brymer, Robert A},
  journal={International Journal of Hospitality Management},
  volume={55},
  pages={41--51},
  year={2016},
  publisher={Elsevier}
}

@article{zou2023social,
  title={Social media for emergency rescue: An analysis of rescue requests on Twitter during Hurricane Harvey},
  author={Zou, Lei and Liao, Danqing and Lam, Nina SN and Meyer, Michelle A and Gharaibeh, Nasir G and Cai, Heng and Zhou, Bing and Li, Dongying},
  journal={International Journal of Disaster Risk Reduction},
  volume={85},
  pages={103513},
  year={2023},
  publisher={Elsevier}
}

@inproceedings{hasan2013understanding,
  title={Understanding urban human activity and mobility patterns using large-scale location-based data from online social media},
  author={Hasan, Samiul and Zhan, Xianyuan and Ukkusuri, Satish V},
  booktitle={Proceedings of the 2nd ACM SIGKDD international workshop on urban computing},
  pages={1--8},
  year={2013}
}

@inproceedings{saha2019social,
  title={Social media as a passive sensor in longitudinal studies of human behavior and wellbeing},
  author={Saha, Koustuv and Bayraktaroglu, Ayse E and Campbell, Andrew T and Chawla, Nitesh V and De Choudhury, Munmun and D'Mello, Sidney K and Dey, Anind K and Gao, Ge and Gregg, Julie M and Jagannath, Krithika and others},
  booktitle={Extended Abstracts of the 2019 CHI Conference on Human Factors in Computing Systems},
  pages={1--8},
  year={2019}
}

@inproceedings{hutto2014vader,
  title={Vader: A parsimonious rule-based model for sentiment analysis of social media text},
  author={Hutto, Clayton and Gilbert, Eric},
  booktitle={Proceedings of the international AAAI conference on web and social media},
  volume={8},
  number={1},
  pages={216--225},
  year={2014}
}

@article{borg2020using,
  title={Using VADER sentiment and SVM for predicting customer response sentiment},
  author={Borg, Anton and Boldt, Martin},
  journal={Expert Systems with Applications},
  volume={162},
  pages={113746},
  year={2020},
  publisher={Elsevier}
}

@article{kaliyar2021fakebert,
  title={FakeBERT: Fake news detection in social media with a BERT-based deep learning approach},
  author={Kaliyar, Rohit Kumar and Goswami, Anurag and Narang, Pratik},
  journal={Multimedia tools and applications},
  volume={80},
  number={8},
  pages={11765--11788},
  year={2021},
  publisher={Springer}
}

@inproceedings{devlin2019bert,
  title={Bert: Pre-training of deep bidirectional transformers for language understanding},
  author={Devlin, Jacob and Chang, Ming-Wei and Lee, Kenton and Toutanova, Kristina},
  booktitle={Proceedings of the 2019 conference of the North American chapter of the association for computational linguistics: human language technologies, volume 1 (long and short papers)},
  pages={4171--4186},
  year={2019}
}

@article{hu2023geo,
  title={Geo-knowledge-guided GPT models improve the extraction of location descriptions from disaster-related social media messages},
  author={Hu, Yingjie and Mai, Gengchen and Cundy, Chris and Choi, Kristy and Lao, Ni and Liu, Wei and Lakhanpal, Gaurish and Zhou, Ryan Zhenqi and Joseph, Kenneth},
  journal={International Journal of Geographical Information Science},
  volume={37},
  number={11},
  pages={2289--2318},
  year={2023},
  publisher={Taylor \& Francis}
}

@inproceedings{lukito2024comparing,
  title={Comparing a BERT classifier and a GPT classifier for detecting connective language across multiple social media},
  author={Lukito, Josephine and Chen, Bin and Masullo, Gina M and Stroud, Natalie Jomini},
  booktitle={Proceedings of the 2024 Conference on Empirical Methods in Natural Language Processing},
  pages={19140--19153},
  year={2024}
}

@inproceedings{duan2025crowdsourcing,
  title={Crowdsourcing-based knowledge graph construction for drug side effects using large language models with an application on semaglutide},
  author={Duan, Zhijie and Wei, Kai and Xue, Zhaoqian and Zhou, Jiayan and Yang, Shu and Ma, Siyuan and Jin, Jin and Li, Lingyao},
  booktitle={AMIA Annual Symposium Proceedings},
  volume={2024},
  pages={332},
  year={2025}
}

@inproceedings{pereira2023crisis,
  title={Crisis event social media summarization with GPT-3 and neural reranking},
  author={Pereira, Jayr and Fidalgo, Robson and Lotufo, Roberto and Nogueira, Rodrigo},
  booktitle={Proceedings of the 20th International ISCRAM Conference},
  pages={371--384},
  year={2023}
}

@article{li2025pixels,
  title={From pixels to places: A systematic benchmark for evaluating image geolocalization ability in large language models},
  author={Li, Lingyao and Yu, Runlong and Hu, Qikai and Li, Bowei and Deng, Min and Zhou, Yang and Jia, Xiaowei},
  journal={arXiv preprint arXiv:2508.01608},
  year={2025}
}

@article{milani2020visual,
  title={The visual vaccine debate on Twitter: A social network analysis},
  author={Milani, Elena and Weitkamp, Emma and Webb, Peter},
  journal={Media and Communication},
  volume={8},
  number={2},
  pages={364--375},
  year={2020}
}

@article{cheng2018social,
  title={How social media is changing crisis communication strategies: Evidence from the updated literature},
  author={Cheng, Yang},
  journal={Journal of contingencies and crisis management},
  volume={26},
  number={1},
  pages={58--68},
  year={2018},
  publisher={Wiley Online Library}
}

@article{al2021public,
  title={Public discourse against masks in the COVID-19 era: Infodemiology study of Twitter data},
  author={Al-Ramahi, Mohammad and Elnoshokaty, Ahmed and El-Gayar, Omar and Nasralah, Tareq and Wahbeh, Abdullah},
  journal={JMIR Public Health and Surveillance},
  volume={7},
  number={4},
  pages={e26780},
  year={2021},
  publisher={JMIR Publications Toronto, Canada}
}

@inproceedings{gauthier2022will,
  title={“I will not drink with you today”: a topic-guided thematic analysis of addiction recovery on Reddit},
  author={Gauthier, Robert P and Costello, Mary Jean and Wallace, James R},
  booktitle={Proceedings of the 2022 CHI Conference on Human Factors in Computing Systems},
  pages={1--17},
  year={2022}
}

@article{lappas2022harnessing,
  title={Harnessing the power of dialogue: examining the impact of facebook content on citizens’ engagement},
  author={Lappas, Georgios and Triantafillidou, Amalia and Kani, Anastasia},
  journal={Local Government Studies},
  volume={48},
  number={1},
  pages={87--106},
  year={2022},
  publisher={Taylor \& Francis}
}

@article{song2022analyze,
  title={Analyze the usage of urban greenways through social media images and computer vision},
  author={Song, Yang and Ning, Huan and Ye, Xinyue and Chandana, Divya and Wang, Shaohua},
  journal={Environment and Planning B: Urban Analytics and City Science},
  volume={49},
  number={6},
  pages={1682--1696},
  year={2022},
  publisher={SAGE Publications Sage UK: London, England}
}

@article{zhang2023can,
  title={Can consumer-posted photos serve as a leading indicator of restaurant survival? Evidence from Yelp},
  author={Zhang, Mengxia and Luo, Lan},
  journal={Management Science},
  volume={69},
  number={1},
  pages={25--50},
  year={2023},
  publisher={INFORMS}
}

@article{treen2022discussion,
  title={Discussion of climate change on Reddit: Polarized discourse or deliberative debate?},
  author={Treen, Kathie and Williams, Hywel and O’Neill, Saffron and Coan, Travis G},
  journal={Environmental Communication},
  volume={16},
  number={5},
  pages={680--698},
  year={2022},
  publisher={Taylor \& Francis}
}

@article{mohamed2024users,
  title={Users’ experience with health-related content on YouTube: an exploratory study},
  author={Mohamed, Fatma and Shoufan, Abdulhadi},
  journal={BMC Public Health},
  volume={24},
  number={1},
  pages={86},
  year={2024},
  publisher={Springer}
}

@article{xu2022analysing,
  title={Analysing information diffusion in natural hazards using retweets-a case study of 2018 winter storm diego},
  author={Xu, Jinwen and Qiang, Yi},
  journal={Annals of GIS},
  volume={28},
  number={2},
  pages={213--227},
  year={2022},
  publisher={Taylor \& Francis}
}

@article{hemphill2021drives,
  title={What drives US congressional members’ policy attention on Twitter?},
  author={Hemphill, Libby and Russell, Annelise and Sch{\"o}pke-Gonzalez, Angela M},
  journal={Policy \& Internet},
  volume={13},
  number={2},
  pages={233--256},
  year={2021},
  publisher={Wiley Online Library}
}

@article{salton1988term,
  title={Term-weighting approaches in automatic text retrieval},
  author={Salton, Gerard and Buckley, Christopher},
  journal={Information processing \& management},
  volume={24},
  number={5},
  pages={513--523},
  year={1988},
  publisher={Elsevier}
}

@inproceedings{manning2014stanford,
  title={The Stanford CoreNLP natural language processing toolkit},
  author={Manning, Christopher D and Surdeanu, Mihai and Bauer, John and Finkel, Jenny Rose and Bethard, Steven and McClosky, David},
  booktitle={Proceedings of 52nd annual meeting of the association for computational linguistics: system demonstrations},
  pages={55--60},
  year={2014}
}

@inproceedings{camacho2022tweetnlp,
  title={TweetNLP: Cutting-edge natural language processing for social media},
  author={Camacho-Collados, Jose and Rezaee, Kiamehr and Riahi, Talayeh and Ushio, Asahi and Loureiro, Daniel and Antypas, Dimosthenis and Boisson, Joanne and Anke, Luis Espinosa and Liu, Fangyu and Mart{\'\i}nez-C{\'a}mara, Eugenio},
  booktitle={Proceedings of the 2022 conference on empirical methods in natural language processing: system demonstrations},
  pages={38--49},
  year={2022}
}

@article{lyu2025gpt,
  title={Gpt-4v (ision) as a social media analysis engine},
  author={Lyu, Hanjia and Huang, Jinfa and Zhang, Daoan and Yu, Yongsheng and Mou, Xinyi and Pan, Jinsheng and Yang, Zhengyuan and Wei, Zhongyu and Luo, Jiebo},
  journal={ACM Transactions on Intelligent Systems and Technology},
  volume={16},
  number={3},
  pages={1--54},
  year={2025},
  publisher={ACM New York, NY}
}

@article{mikolov2013distributed,
  title={Distributed representations of words and phrases and their compositionality},
  author={Mikolov, Tomas and Sutskever, Ilya and Chen, Kai and Corrado, Greg S and Dean, Jeff},
  journal={Advances in neural information processing systems},
  volume={26},
  year={2013}
}

@inproceedings{pennington2014glove,
  title={Glove: Global vectors for word representation},
  author={Pennington, Jeffrey and Socher, Richard and Manning, Christopher D},
  booktitle={Proceedings of the 2014 conference on empirical methods in natural language processing (EMNLP)},
  pages={1532--1543},
  year={2014}
}

@article{bojanowski2017enriching,
  title={Enriching word vectors with subword information},
  author={Bojanowski, Piotr and Grave, Edouard and Joulin, Armand and Mikolov, Tomas},
  journal={Transactions of the association for computational linguistics},
  volume={5},
  pages={135--146},
  year={2017},
  publisher={MIT Press One Rogers Street, Cambridge, MA 02142-1209, USA journals-info~…}
}

@article{liu2019roberta,
  title={Roberta: A robustly optimized bert pretraining approach},
  author={Liu, Yinhan and Ott, Myle and Goyal, Naman and Du, Jingfei and Joshi, Mandar and Chen, Danqi and Levy, Omer and Lewis, Mike and Zettlemoyer, Luke and Stoyanov, Veselin},
  journal={arXiv preprint arXiv:1907.11692},
  year={2019}
}

@article{brown2020language,
  title={Language models are few-shot learners},
  author={Brown, Tom and Mann, Benjamin and Ryder, Nick and Subbiah, Melanie and Kaplan, Jared D and Dhariwal, Prafulla and Neelakantan, Arvind and Shyam, Pranav and Sastry, Girish and Askell, Amanda and others},
  journal={Advances in neural information processing systems},
  volume={33},
  pages={1877--1901},
  year={2020}
}

@article{hu2022lora,
  title={Lora: Low-rank adaptation of large language models.},
  author={Hu, Edward J and Shen, Yelong and Wallis, Phillip and Allen-Zhu, Zeyuan and Li, Yuanzhi and Wang, Shean and Wang, Liang and Chen, Weizhu and others},
  journal={Iclr},
  volume={1},
  number={2},
  pages={3},
  year={2022}
}

@article{gao2019target,
  title={Target-dependent sentiment classification with BERT},
  author={Gao, Zhengjie and Feng, Ao and Song, Xinyu and Wu, Xi},
  journal={Ieee Access},
  volume={7},
  pages={154290--154299},
  year={2019},
  publisher={IEEE}
}

@inproceedings{barbieri2020tweeteval,
  title={TweetEval: Unified benchmark and comparative evaluation for tweet classification},
  author={Barbieri, Francesco and Camacho-Collados, Jose and Anke, Luis Espinosa and Neves, Leonardo},
  booktitle={Findings of the association for computational linguistics: EMNLP 2020},
  pages={1644--1650},
  year={2020}
}

@inproceedings{nadi2023sentiment,
  title={Sentiment analysis using large language models: A case study of GPT-3.5},
  author={Nadi, Farhad and Naghavipour, Hadi and Mehmood, Tahir and Azman, Alliesya Binti and Nagantheran, Jeetha A/P and Ting, Kezia Sim Kui and Adnan, Nor Muhammad Ilman Bin Nor and Sivarajan, Roshene A/P and Veerah, Suita A/P and Rahmat, Romi Fadillah},
  booktitle={The International Conference on Data Science and Emerging Technologies},
  pages={161--168},
  year={2023},
  organization={Springer}
}

@article{kuccuk2020stance,
  title={Stance detection: A survey},
  author={K{\"u}{\c{c}}{\"u}k, Dilek and Can, Fazli},
  journal={ACM Computing Surveys (CSUR)},
  volume={53},
  number={1},
  pages={1--37},
  year={2020},
  publisher={ACM New York, NY, USA}
}

@article{karande2021stance,
  title={Stance detection with BERT embeddings for credibility analysis of information on social media},
  author={Karande, Hema and Walambe, Rahee and Benjamin, Victor and Kotecha, Ketan and Raghu, TS},
  journal={PeerJ Computer Science},
  volume={7},
  pages={e467},
  year={2021},
  publisher={PeerJ Inc.}
}

@article{ziems2024can,
  title={Can large language models transform computational social science?},
  author={Ziems, Caleb and Held, William and Shaikh, Omar and Chen, Jiaao and Zhang, Zhehao and Yang, Diyi},
  journal={Computational Linguistics},
  volume={50},
  number={1},
  pages={237--291},
  year={2024}
}

@inproceedings{zhang2024sentiment,
  title={Sentiment analysis in the era of large language models: A reality check},
  author={Zhang, Wenxuan and Deng, Yue and Liu, Bing and Pan, Sinno and Bing, Lidong},
  booktitle={Findings of the Association for Computational Linguistics: NAACL 2024},
  pages={3881--3906},
  year={2024}
}

@article{mughal2024comparative,
  title={Comparative analysis of deep natural networks and large language models for aspect-based sentiment analysis},
  author={Mughal, Nimra and Mujtaba, Ghulam and Shaikh, Sarang and Kumar, Aveenash and Daudpota, Sher Muhammad},
  journal={Ieee Access},
  volume={12},
  pages={60943--60959},
  year={2024},
  publisher={IEEE}
}

@article{jelodar2019latent,
  title={Latent Dirichlet allocation (LDA) and topic modeling: models, applications, a survey},
  author={Jelodar, Hamed and Wang, Yongli and Yuan, Chi and Feng, Xia and Jiang, Xiahui and Li, Yanchao and Zhao, Liang},
  journal={Multimedia tools and applications},
  volume={78},
  number={11},
  pages={15169--15211},
  year={2019},
  publisher={Springer}
}

@article{grootendorst2022bertopic,
  title={BERTopic: Neural topic modeling with a class-based TF-IDF procedure},
  author={Grootendorst, Maarten},
  journal={arXiv preprint arXiv:2203.05794},
  year={2022}
}

@article{mcinnes2018umap,
  title={Umap: Uniform manifold approximation and projection for dimension reduction},
  author={McInnes, Leland and Healy, John and Melville, James},
  journal={arXiv preprint arXiv:1802.03426},
  year={2018}
}

@inproceedings{pham2024topicgpt,
  title={TopicGPT: A prompt-based topic modeling framework},
  author={Pham, Chau Minh and Hoyle, Alexander and Sun, Simeng and Resnik, Philip and Iyyer, Mohit},
  booktitle={Proceedings of the 2024 Conference of the North American Chapter of the Association for Computational Linguistics: Human Language Technologies (Volume 1: Long Papers)},
  pages={2956--2984},
  year={2024}
}

@article{li2020survey,
  title={A survey on deep learning for named entity recognition},
  author={Li, Jing and Sun, Aixin and Han, Jianglei and Li, Chenliang},
  journal={IEEE transactions on knowledge and data engineering},
  volume={34},
  number={1},
  pages={50--70},
  year={2020},
  publisher={IEEE}
}

@article{li2022has,
  title={How has airport service quality changed in the context of COVID-19: A data-driven crowdsourcing approach based on sentiment analysis},
  author={Li, Lingyao and Mao, Yujie and Wang, Yu and Ma, Zihui},
  journal={Journal of Air Transport Management},
  volume={105},
  pages={102298},
  year={2022},
  publisher={Elsevier}
}

@article{li2020leveraging,
  title={Leveraging social media data to study the community resilience of New York City to 2019 power outage},
  author={Li, Lingyao and Ma, Zihui and Cao, Tao},
  journal={International Journal of Disaster Risk Reduction},
  volume={51},
  pages={101776},
  year={2020},
  publisher={Elsevier}
}

@article{hemati2019lstmvoter,
  title={LSTMVoter: chemical named entity recognition using a conglomerate of sequence labeling tools},
  author={Hemati, Wahed and Mehler, Alexander},
  journal={Journal of cheminformatics},
  volume={11},
  number={1},
  pages={3},
  year={2019},
  publisher={Springer}
}

@article{li2025llm,
  title={LLM Use for Mental Health: Crowdsourcing Users' Sentiment-based Perspectives and Values from Social Discussions},
  author={Li, Lingyao and Huang, Xiaoshan and Ma, Renkai and Zhang, Ben Zefeng and Wu, Haolun and Yang, Fan and Chen, Chen},
  journal={arXiv preprint arXiv:2512.07797},
  year={2025}
}

@article{paradkar2022examining,
  title={Examining the consistency between geo-coordinates and content-mentioned locations in tweets for disaster situational awareness: A Hurricane Harvey study},
  author={Paradkar, Aumkar Shriram and Zhang, Cheng and Yuan, Faxi and Mostafavi, Ali},
  journal={International Journal of Disaster Risk Reduction},
  volume={73},
  pages={102878},
  year={2022},
  publisher={Elsevier}
}

@article{honnibal2020spacy,
  title={spaCy: Industrial-strength natural language processing in Python},
  author={Honnibal, Matthew and Montani, Ines and Van Landeghem, Sofie and Boyd, Adriane and others},
  year={2020},
  publisher={Zenodo, Honolulu, HI, USA}
}

@inproceedings{hu2022beyond,
  title={Beyond geo-localization: Fine-grained orientation of street-view images by cross-view matching with satellite imagery},
  author={Hu, Wenmiao and Zhang, Yichen and Liang, Yuxuan and Yin, Yifang and Georgescu, Andrei and Tran, An and Kruppa, Hannes and Ng, See-Kiong and Zimmermann, Roger},
  booktitle={Proceedings of the 30th ACM international conference on multimedia},
  pages={6155--6164},
  year={2022}
}

@article{schober2018correlation,
  title={Correlation coefficients: appropriate use and interpretation},
  author={Schober, Patrick and Boer, Christa and Schwarte, Lothar A},
  journal={Anesthesia \& analgesia},
  volume={126},
  number={5},
  pages={1763--1768},
  year={2018},
  publisher={LWW}
}

@book{draper1998applied,
  title={Applied regression analysis},
  author={Draper, Norman R and Smith, Harry},
  volume={326},
  year={1998},
  publisher={John Wiley \& Sons}
}

@article{fotheringham2017multiscale,
  title={Multiscale geographically weighted regression (MGWR)},
  author={Fotheringham, A Stewart and Yang, Wenbai and Kang, Wei},
  journal={Annals of the American Association of Geographers},
  volume={107},
  number={6},
  pages={1247--1265},
  year={2017},
  publisher={Taylor \& Francis}
}

@article{anselin2022spatial,
  title={Spatial econometrics},
  author={Anselin, Luc},
  journal={Handbook of spatial analysis in the social sciences},
  pages={101--122},
  year={2022},
  publisher={Edward Elgar Publishing}
}

@misc{census_hps,
  author={{U.S. Census Bureau}},
  title={Household Pulse Survey},
  year={2026},
  howpublished={\url{https://www.census.gov/programs-surveys/household-pulse-survey.html}},
  note={Accessed: 2026-04-08}
}

@misc{census_acs,
  author={{U.S. Census Bureau}},
  title={American Community Survey (ACS)},
  year={2026},
  howpublished={\url{https://www.census.gov/programs-surveys/acs.html}},
  note={Accessed: 2026-04-08}
}

@article{wu2018disaster,
  title={Disaster early warning and damage assessment analysis using social media data and geo-location information},
  author={Wu, Desheng and Cui, Yiwen},
  journal={Decision support systems},
  volume={111},
  pages={48--59},
  year={2018},
  publisher={Elsevier}
}

@article{kostkova2014swineflu,
  title={\# swineflu: The use of twitter as an early warning and risk communication tool in the 2009 swine flu pandemic},
  author={Kostkova, Patty and Szomszor, Martin and St. Louis, Connie},
  journal={ACM Transactions on Management Information Systems (TMIS)},
  volume={5},
  number={2},
  pages={1--25},
  year={2014},
  publisher={ACM New York, NY, USA}
}

@incollection{ma2023appraising,
  title={Appraising Situational Awareness in Social Media Data for Wildfire Response},
  author={Ma, Zihui and Li, Lingyao and Yuan, Yubai and Baecher, Gregory B},
  booktitle={ASCE Inspire 2023},
  pages={289--297},
  year={2023}
}

@article{blank2017representativeness,
  title={Representativeness of social media in great britain: investigating Facebook, Linkedin, Twitter, Pinterest, Google+, and Instagram},
  author={Blank, Grant and Lutz, Christoph},
  journal={American Behavioral Scientist},
  volume={61},
  number={7},
  pages={741--756},
  year={2017},
  publisher={Sage Publications Sage CA: Los Angeles, CA}
}

@inproceedings{yin2018evaluating,
  title={Evaluating the representativeness in the geographic distribution of twitter user population},
  author={Yin, Junjun and Chi, Guangqing and Van Hook, Jennifer},
  booktitle={Proceedings of the 12th workshop on geographic information retrieval},
  pages={1--2},
  year={2018}
}

@inproceedings{zhuo2024prosa,
  title={ProSA: Assessing and understanding the prompt sensitivity of LLMs},
  author={Zhuo, Jingming and Zhang, Songyang and Fang, Xinyu and Duan, Haodong and Lin, Dahua and Chen, Kai},
  booktitle={Findings of the Association for Computational Linguistics: EMNLP 2024},
  pages={1950--1976},
  year={2024}
}

@inproceedings{li2023halueval,
  title={Halueval: A large-scale hallucination evaluation benchmark for large language models},
  author={Li, Junyi and Cheng, Xiaoxue and Zhao, Wayne Xin and Nie, Jian-Yun and Wen, Ji-Rong},
  booktitle={Proceedings of the 2023 conference on empirical methods in natural language processing},
  pages={6449--6464},
  year={2023}
}

@article{dang2023x_api,
  author  = {Dang, Sheila},
  title   = {Exclusive: Elon Musk's X restructuring curtails disinformation research, spurs legal fears},
  journal = {Reuters},
  year    = {2023},
  month   = nov,
  day     = {6},
  url      = {https://www.reuters.com/technology/elon-musks-x-restructuring-curtails-disinformation-research-spurs-legal-fears-2023-11-06/},
  note    = {Accessed: 2026-04-08}
}
\end{document}